\documentclass[aps,prd,reprint,twocolumn,superscriptaddress,nofootinbib]{revtex4-2}
\usepackage[utf8]{inputenc}
\usepackage[english]{babel}
\usepackage{amsmath}
\usepackage{amsfonts}
\usepackage{amssymb}
\usepackage{graphicx}
\usepackage{bm}
\usepackage{float}

\begin{document}

\title{Wave propagation and hybridization of plasmonic modes in Maxwell-Chern-Simons pseudo-electrodynamics}

\author{S. Duque Cesar}
\email{samduquec@gmail.com}
\affiliation{Departamento de Física, Universidade Federal Rural do Rio de Janeiro, 
\\
BR 465-07, 23890-971, Seropédica, RJ, Brasil}

\author{M. J. Neves}
\email{mariojr@ufrrj.br}
\affiliation{Departamento de Física, Universidade Federal Rural do Rio de Janeiro, 
\\
BR 465-07, 23890-971, Seropédica, RJ, Brasil}

\begin{abstract}

We investigate the propagation of classical plane waves and surface plasmon-polariton modes within the framework of pseudo-electrodynamics (PED) supplemented by a non-local Chern-Simons (CS) topological term. Starting from the dimensionally reduced action, we derive the decoupled second-order wave equations for the electromagnetic field, and for the gauge-potential. In vacuum, we show that the topological CS parameter works as a dynamical mass generator, modifying the strict transversality of plane waves, and introducing a distinct energy \textit{gap} in the dispersion relation. Considering a planar conducting medium that satisfies the Ohm law, the topological mass induces a novel hybridization mechanism between the Transverse Electric (TE), and Transverse Magnetic (TM) modes, a feature entirely absent in the conventional planar plasmonics. We analyze the asymptotic limits of this system, obtaining the exact real solutions in the lossless reactive Drude regime, and deriving the complex refractive index through a quasi-local approximation in the dissipative regime.

\end{abstract}

\maketitle

\section{Introduction}
The connection between quantum field theories in low dimensions and condensed matter physics has gained relevance in recent decades due to applications in planar materials \cite{ando1975, tsui1982, laughlin1983, jablan2009, grigorenko2012}. Planar materials such as graphene, Weyl semimetals, and topological insulators exhibit exotic transport properties that are successfully modeled by a gauge theory configurations in the reduced space-time \cite{marino2017}. Since the charge carriers are strictly confined to a two-dimensional spatial surface, the electromagnetic fields that mediate their interactions continue to propagate freely throughout the surrounding three-dimensional space.
The description of the interaction is so introduced through a dimensional reduction of Maxwell's electrodynamics, projecting the \textit{bulk} interaction on a bidimensional plane \cite{marino1993, amaral1992}. This procedure results in an effective field theory known as Pseudo-Electrodynamics (PED), which is intrinsically non-local, and characterized by infinite-order derivatives in the d'Alembertian operator $2/\sqrt{-\bar\Box}$. Despite its non-local architecture, it has been demonstrated that PED preserves fundamental principles such as the gauge invariance, unitarity, and causality \cite{marino2014, alves2019, magalhaes2020,neves2025,nevesPLB}.
In a recent work, we investigated the classical aspects of pseudo-electrodynamics \cite{duque2025}, focusing on the conservation laws, the construction of the symmetric and gauge-invariant Belinfante-Rosenfeld energy-momentum tensor, and the radiation fields produced by accelerated point charges. However, a comprehensive analysis of vacuum excitations, wave kinematics, and the field response to material boundaries remains unexplored. Furthermore, the inclusion of a non-local counterpart to the Chern-Simons topological term provides a mechanism for topological mass generation, which profoundly alters the behavior of propagating perturbations in material physics.
In this letter, we use the Maxwell-Chern-Simons pseudo-electrodynamics (PED-MCS) to study the wave propagation, the dispersion relations, and surface plasmon excitations, when the planar material is governed by the Ohm law. We obtain the refractive index for plane wave solutions that propagate through the planar materials in the presence of electric condutivity, and also of the topological properties due to the Chern-Simons parameter.    
The letter is organized as follows : In section II, we derive the general wave equations for the EM field and the gauge potential, solving these equations for plane waves in the vacuum, and analyzing the topological mass (\textit{gap}). The section III explores wave propagation within an Ohmic planar conductor, demonstrating the hybridization of TE- and TM-modes through the reactive and dissipative regimes. Finally, our conclusions and future perspectives are presented in the section IV.
We use the natural units of $\hbar=c=1$ in which the space-time in $1+2$ dimensions is set by the Minkowski metric $\eta^{\bar{\mu}\bar{\nu}}=\mbox{diag}(+1,-1,-1)$, and the bar greek index runs as $\bar{\mu} \, , \, \bar{\nu}=\{\, 0,1,2 \, \}$ for vectors and tensors.

\section{Wave equations in topological pseudo-electrodynamics}
The dynamics of non-local Maxwell-Chern-Simons pseudo-electrodynamics in $1+2$ dimensions is governed by the Lagrangian density \cite{MJNevesEPJC2026} :
\begin{equation}\label{LPEDMCS}
\begin{split}
    \mathcal{L}_{PEDMCS} &= -\frac{1}{4}\,F_{\bar{\mu}\bar{\nu}}\left[\frac{2}{\sqrt{-\bar\Box}}F^{\bar{\mu}\bar{\nu}}\right] \\
    &\quad + \frac{\theta}{2}\epsilon^{\bar{\mu}\bar{\nu}\bar{\rho}}A_{\bar{\mu}}\left[\frac{2}{\sqrt{-\bar\Box}}\partial_{\bar{\nu}}A_{\bar{\rho}}\right] - j_{\bar{\mu}}A^{\bar{\mu}} \; ,
\end{split}
\end{equation}
where $F_{\bar{\mu}\bar{\nu}}=\partial_{\bar{\mu}}A_{\bar{\nu}}-\partial_{\bar{\nu}}A_{\bar{\mu}}=(E^{i},-B)$ is the electromagnetic field strength tensor, $\theta$ is the topological mass parameter, $\bar\Box = \partial_t^2 - \partial_{x}^2-\partial_{y}^2$ is the d'Alembertian operator in $1+2$ dimensions, and $j^{\bar{\mu}}$ is a classical source constrained on a planar space. The bar greek index emphasize that they run over the $1+2$ space-time $\bar{\mu},\bar{\nu}=\{ \,0,1,2 \,\}$. The action associated with the lagrangian (\ref{LPEDMCS}) is invariant under gauge transformations in the $A^{\bar{\mu}}$-gauge potential. Applying the action principle yields the covariant field equations with sources :
\begin{equation}
    \frac{2}{\sqrt{-\bar\Box}}(\partial^{\bar{\mu}}F_{\bar{\mu}\bar{\nu}}) + \theta\frac{2}{\sqrt{-\bar\Box}}\tilde{F}_{\bar{\nu}} = j_{\bar{\nu}} \; ,
    \label{fieldeq}
\end{equation}
in which the field equations are completed by the Bianchi identity $\partial_{\bar{\mu}}\tilde{F}^{\bar{\mu}}=0$, where $\tilde{F}^{\bar{\mu}} = \frac{1}{2}\,\epsilon^{\bar{\mu}\bar{\nu}\bar{\rho}}F_{\bar{\nu}\bar{\rho}} = \left( \, B \, , \, \epsilon^{ij}\,E^j \right)$ is the dual strength field vector.
The CS term does the previous equations to be coupled. To obtain wave equations for the EM field, we operate on the Bianchi identity with the non-local operator and spatial derivatives. Using Eq.~(\ref{fieldeq}), we find that the dual field satisfies a Klein-Gordon-type dynamics:
\begin{equation}
    \frac{2}{\sqrt{-\bar\Box}}\left(\bar\Box + \theta^2\right)\tilde{F}_{\bar{\alpha}} = \theta \, j_{\bar{\alpha}} + \epsilon_{\bar{\mu}\bar{\nu}\bar{\alpha}}\,\partial^{\bar{\mu}}j^{\bar{\nu}} \; .
\end{equation}
The decomposition of this expression into temporal ($\bar{\alpha}=0$) and spatial ($\bar{\alpha}=i$) components yields 
the exact wave equations for the observable pseudo-scalar magnetic field $B$, and for the electric field $\mathbf{E}$ :
\begin{subequations}
\begin{eqnarray}
    \frac{2}{\sqrt{-\bar\Box}}\left(\bar\Box + \theta^2\right)B &=& \theta \sigma + \partial_x j_y - \partial_y j_x \; ,
\\    
    \frac{2}{\sqrt{-\bar\Box}} \left(\bar\Box + \theta^2\right)E_x &=& -\theta j_y - \partial_t j_x + \partial_x \sigma \; ,
\\
    \frac{2}{\sqrt{-\bar\Box}} \left(\bar\Box + \theta^2\right)E_y &=& \theta j_x - \partial_t j_y + \partial_y \sigma \; ,
\end{eqnarray}
\end{subequations}
where $\sigma = j_0$ is the surface charge density. Similarly, choosing the Lorenz gauge condition $\partial_{\bar{\mu}}A^{\bar{\mu}}=0$, the wave equation for the $A_{\bar{\rho}}$-potential is given by
\begin{equation}
\frac{2}{\sqrt{-\bar\Box}} \left(\bar\Box + \theta^2\right) A_{\bar{\rho}} = j_{\bar{\rho}} 
- \theta \, \epsilon_{\bar{\rho}\bar{\alpha}\bar{\beta}} \, \frac{1}{\bar\Box}\partial^{\bar{\alpha}}j^{\bar{\beta}} \; .
\end{equation}
These equations show that fields and potentials propagate under a massive differential structure, governed by the operator $(\bar\Box + \theta^2)$.

To probe the fundamental excitations in the vacuum, we consider free propagation in the absence of external sources ($\sigma = 0, \mathbf{j} = 0$). We seek plane wave solutions of the form :
\begin{equation}
\mathbf{E}(\mathbf{r},t) = \mathbf{E}_0  \, e^{i(\mathbf{k}\cdot\mathbf{r}-\omega t)} , \quad B(\mathbf{r},t) = B_0 \, e^{i(\mathbf{k}\cdot\mathbf{r}-\omega t)} \; ,
\end{equation}
where $({\bf E}_{0},B_{0})$ are the electric and magnetic amplitudes (constants and uniform), $\mathbf{k} = (k_x, k_y)$ is the wave vector, and $\omega$ is the frequency. Under these solutions, the non-local operator acts by mapping the eigenvalues, extracting the expression $\frac{2}{\sqrt{-\bar\Box}} \rightarrow \frac{2}{\sqrt{\omega^2 - {\bf k}^2}}$ in the frequency space. Substituting these wave solutions into the source-free field equations (modified Gauss, Ampère-Maxwell, and Faraday laws), we obtain the following coupled algebraic system for the electric and magnetic amplitudes :
\begin{subequations}
\label{eq:sistema_livre}
\begin{align}
    \mathbf{k}\cdot\mathbf{E}_0 &= -\,i\,\theta \, B_0 \; , \label{eq:gauss_mod} \\
    \omega \, \mathbf{E}_{0j} &= \epsilon_{ij} \, k_i \, B_0 + i \, \theta \, \epsilon_{ji} \, E_{0i} \; , 
    \label{eq:ampere_mod} \\
    \omega \, B_0 &= \epsilon_{lm} \, k_l \, E_{0m} \; . \label{eq:faraday_mod}
\end{align}
\end{subequations}
The equation (\ref{eq:gauss_mod}) shows that the presence of the topological term breaks the strict transversality of conventional electrodynamics \cite{jackson1999}. The electric field develops a longitudinal component parallel to $\mathbf{k}$ that oscillates with a phase shift of $\pi/2$ relative to the magnetic field. This longitudinal degree of freedom is the explicit manifestation of a massive vector boson in $1+2$ dimensions.
The combination of Eqs.~(\ref{eq:ampere_mod}) and (\ref{eq:faraday_mod}) leads to the wave equation for the electric amplitude, $\mathbf{M}_{ji}\mathbf{E}_{0i}=0$, in which the wave matrix is 
\begin{equation}
\mathbf{M}_{ji} = \left(\omega^2 - k^2\right)\delta_{ij} + k_i \, k_j - i \, \omega \, \theta \, \epsilon_{ji} \; . 
\end{equation}
Requiring a non-trivial solution demands that $\det(\mathbf{M})=\omega^2(\omega^2 - {\bf k}^2 - \theta^2)=0$, which leads to dispersion relation :
\begin{equation}
\omega({\bf k}) = \sqrt{ {\bf k}^2 + \theta^2} \; .
\end{equation}
The result confirms that the non-local Chern-Simons term establishes a clean energy \textit{gap} proportional to $\theta$. The dispersion relation is identical to that of a massive free relativistic particle, showing that topological mass generation is achieved natively without the need to add scalar fields.
\section{Wave propagation in planar conductors and plasmonic hybridization}

We now expand our study to consider a realistic two-dimensional conducting medium ({\it e.g.}, dopped graphene or a thin metallic film) characterized by an electric conductivity $\sigma_c$. The induced current is governed by the local Ohm's law:
\begin{equation}
\mathbf{j} = \sigma_c \, \mathbf{E} \; .
\end{equation}
Since that electric charge is conserved, the sources satisfy the continuity equation $\partial_t \rho + \bar{\nabla}\cdot\mathbf{j} = 0$. 
For the plane wave solution, the charge conservation leads to relation $\rho = \frac{\sigma_c}{\omega}(\mathbf{k}\cdot\mathbf{E})$. Substituting this response into the modified field equations, the competition between non-locality, topology, and material response 
yields the general equation for the modes :
\begin{equation}
\left( \omega^2 - k^2 + \frac{i\omega\sigma_c}{\mathcal{R}} \right) \mathbf{E}_{0j} = -k_j \left(\mathbf{k}\cdot\mathbf{E}_0\right) + i\,\omega\,\theta\,\epsilon_{ji}\,\mathbf{E}_{0i} \; ,
\end{equation}
where $\mathcal{R} = \frac{2}{\sqrt{\omega^2 - k^2}}$. Considering the wave propagation on the ${\cal X}$-direction, $\mathbf{k} = (k,0)$, that maps the problem into a $2\times2$ system that separates the longitudinal component $E_{0x}$ (TM-like) and the transverse component $E_{0y}$ (TE-like) : 
\begin{equation}
    \begin{pmatrix}
        \omega^2 + \frac{i\,\omega\,\sigma_c}{\mathcal{R}} & -i\,\omega\,\theta 
        \\
        \\
        i\,\omega\,\theta & \omega^2 - k^2 + \frac{i\,\omega\,\sigma_c}{\mathcal{R}}
    \end{pmatrix}
    \begin{pmatrix}
        E_{0x} 
        \\
        \\
        E_{0y}
    \end{pmatrix} = 0 \; .
\end{equation}
The null determinant provides the dispersion relation for the topological surface plasmon-polaritons :
\begin{equation}\label{genplasmon}
\left( \omega^2 + \frac{i\,\omega\,\sigma_c}{\mathcal{R}} \right)\left( \omega^2 - k^2 + \frac{i\,\omega\,\sigma_c}{\mathcal{R}} \right) = \omega^2 \, \theta^2 \; .
\end{equation}
In standard planar plasmonics ($\theta=0$), the off-diagonal mixing terms are zero, allowing the pure TM plasmons to propagate independently. However, when $\theta \neq 0$, the topological mass acts as a mixing parameter, induzing a hybridization between the TE- and TM-modes. The planar space-time structure acts on the free electron analogously to a classical Hall Effect, producing a chiral and elliptically polarized mode that is topologically protected against backscattering.
\subsection{The lossless reactive regime}
At high frequencies regime, the material response is dominated by electron inertia rather than scattering. According to the Drude model \cite{ziman1972}, the electric conductivity becomes purely imaginary: $\sigma_c = i\,\sigma_2$ ($\sigma_2 > 0$). Seeking the propagating and subluminal modes ($\omega < k$), the non-local factor becomes imaginary: $\sqrt{\omega^2 - k^2} = i\sqrt{k^2 - \omega^2}$. Under these conditions, the imaginary factors compensate each other, and the Eq.~(\ref{genplasmon}) collapses into a purely real algebraic relation for $\theta = 0$ is $\sqrt{k^2 - \omega^2} = \frac{\sigma_2}{2} \, \omega$, that leads to :
\begin{equation}
\omega(k) = \frac{2k}{\sqrt{\sigma_2^2 + 4}} \; .
\end{equation}
It recovers the usual linear dispersion relation of the conventional 2D plasmons. When $\theta$ is turned on, numerical simulations show that this mode is driven upwards, exhibiting the emergence of a pure massive \textit{Gap} ($\Delta_{gap}$) $\omega_{min} \approx \theta$, when $k \rightarrow 0$, shielding the system against low-frequency excitations. Figure 1 below shows a plot, generated in Python, that compares the dispersion spectrum of the plasmonic modes. The curves indicate the relationship between the frequency and the wave vector with or without the presence of the Chern-Simons term.

\begin{figure}[H]
    \centering
    \includegraphics[width=\columnwidth]{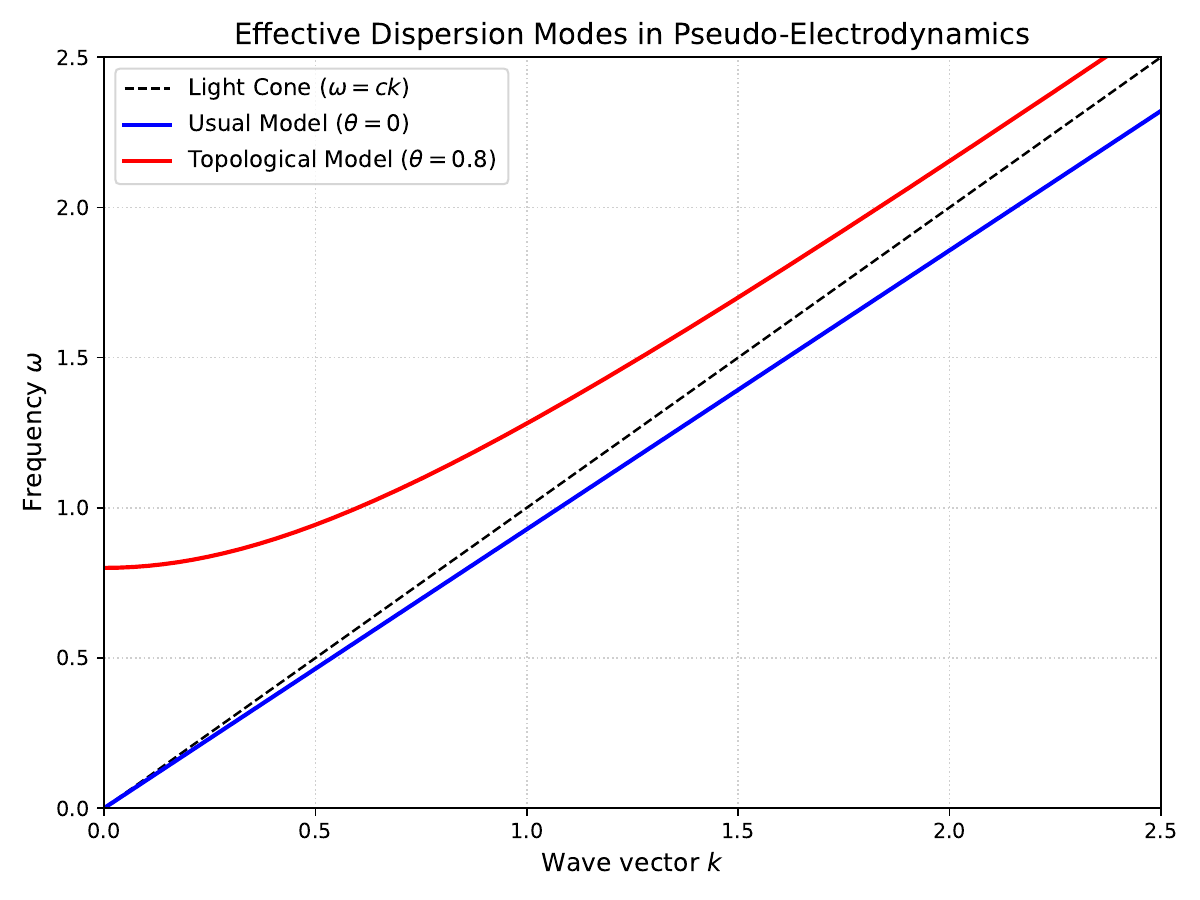}
    \caption{\textbf{Dispersion Spectrum of Plasmonic Modes.} Comparison between the usual model and the massive topological model. The blue curve (Usual, $\theta=0$) exhibits the typical behavior of a gapless surface plasmon. The red curve (Topological, $\theta=0.4$) shows the opening of an energy gap and the alteration of the wave's group velocity. We have used the electrict conductivity : $\sigma_c = 0.8i$.}
\end{figure}
\subsection{The dissipative regime}
In a resistive medium in which the electric conductivity is purely real, the system exhibits Ohmic dissipation, resulting into the decay of the electromagnetic wave. In order to obtain the dispersion relation exactly, avoiding perturbative expansions in the long-wavelength limit ($k \ll \omega$), we parameterize the equation directly in terms of the refractive index $n = k/\omega$. 
From the definition of the non-local operator $\mathcal{R}$, the inverse of $\mathcal{R}$ takes the form :
\begin{equation}
\frac{1}{\mathcal{R}} = \frac{1}{2}\sqrt{\omega^2 - k^2} = \frac{\omega}{2}\sqrt{1 - n^2} \; .
\end{equation}
Introducing the dimensionless complex variable $z = \sqrt{1 - n^2}$, the equation (\ref{genplasmon}) is rewritten as
%
%
%
%
\begin{equation}
    i \, \frac{\sigma_c}{2} \, z^3 + \left( 1 - \frac{\sigma_c^2}{4} \right) z^2 + i \, \frac{\sigma_c}{2} \, z - \frac{\theta^2}{\omega^2} = 0 \; .
\end{equation}
The physical consistency of this polynomial is verified in the limit of null topological mass ($\theta = 0$), in which the equation simplifies as
\begin{equation}
z \left(1 + i\frac{\sigma_c}{2}z\right) \left(z + i\frac{\sigma_c}{2}\right) = 0 \; .
\end{equation}
This limit reproduces the three decoupled solutions of conventional planar electrodynamics : $z=0$, or $n=1$ (the usual vaccum solution), 
$z = -i\sigma_c/2$ (the TM two-dimensional plasmon), and $z = 2i/\sigma_c$ (the damped TE mode). 
For $\theta \neq 0$, the modes undergo hybridization. To extract the analytical root, the polynomial equation is rewritten in its canonical form $z^3 + a_2 \, z^2 + a_1 \, z + a_0 = 0$ by dividing it by $i\,\sigma_c/2$ :
\begin{equation}
a_2 = -i\left(\frac{4 - \sigma_c^2}{2\sigma_c}\right), \quad a_1 = 1, \quad a_0 = i\left(\frac{2\theta^2}{\sigma_c\omega^2}\right) \; .
\end{equation}
The nature of the coupling shows that $a_1$ is strictly real, while $a_2$ and $a_0$ are purely imaginary. Applying the Tschirnhaus transformation \cite{gradshteyn2014}, in which $z = w - \frac{a_2}{3}$, it eliminates the quadratic term, reducing the expression 
to the cubic form $w^3 + pw + q = 0$, with the parameters $p$ and $q$ defined as:
\begin{subequations}
\begin{eqnarray}
p &=& a_1 - \frac{a_2^2}{3} = 1 + \frac{1}{3}\left(\frac{4 - \sigma_c^2}{2\sigma_c}\right)^2 \; ,
\\
q &=& a_0 - \frac{a_1 a_2}{3} + \frac{2a_2^3}{27} 
\nonumber \\
&=& i \left[ \frac{2\theta^2}{\sigma_c \omega^2} + \frac{4 - \sigma_c^2}{6\sigma_c} + \frac{2}{27}\left(\frac{4 - \sigma_c^2}{2\sigma_c}\right)^3 \right] \; .
\end{eqnarray}
\end{subequations}
For any real conductivity $\sigma_c$, it is verified that $p$ is positive, and $q$ is a pure imaginary. The Cardano discriminant 
for this equation is :
\begin{equation}
\Delta = \left(\frac{q}{2}\right)^2 + \left(\frac{p}{3}\right)^3 = -\frac{|q|^2}{4} + \frac{p^3}{27} \; .
\end{equation}
Since $\Delta$ is purely real, it acts as a threshold that distinguishes the phase regimes of the wave (propagating, evanescent, or critically damped). We use the Tschirnhaus transformations through the formula :
\begin{equation}\label{wm}
w_m = \xi^m \sqrt[3]{-\frac{q}{2} + \sqrt{\Delta}} + \xi^{3-m} \sqrt[3]{-\frac{q}{2} - \sqrt{\Delta}} \; ,
\end{equation}
where $m = \{0, 1, 2\}$, and $\xi = e^{i2\pi/3}$ is the primitive cubic root of unity. The existence of three distinct roots reflects the diversity of electromagnetic excitations supported by the planar space-time. The solutions corresponding to $m=1$ and $m=2$ describe, respectively, a radiative bulk mode (associated with scattering and the modified background photon that escapes the boundary) and an overdamped evanescent mode (dominated by Ohmic dissipation and strictly confined to the near-field). 
Since our primary focus is the study of surface plasmon-polaritons—which require both longitudinal wave propagation and finite spatial confinement without violating causality—these two modes are analytically discarded. Therefore, the selection of the proper physical root ($m=0$) is strictly fixed by the thermodynamic condition of spatial energy attenuation and the requirement of dispersion continuity in the limit $\sigma_c \rightarrow 0$. Reverting the transformations, the selected root of the non-local problem is:
\begin{equation}\label{zm}
z_{m} = w_{m} + i\left(\frac{4 - \sigma_c^2}{6\sigma_c}\right) \; .
\end{equation}
Consequently, the complex refractive index of the topological plasmon is determined exactly by 
inverting the initial parameterization $n_{m} = \sqrt{1 - z_{m}^2}$, and using the expressions of (\ref{wm}) and (\ref{zm}), 
we obtain the result
\vspace*{\fill}
\begin{figure*}[t]
    \centering
    \includegraphics[width=0.95\textwidth]{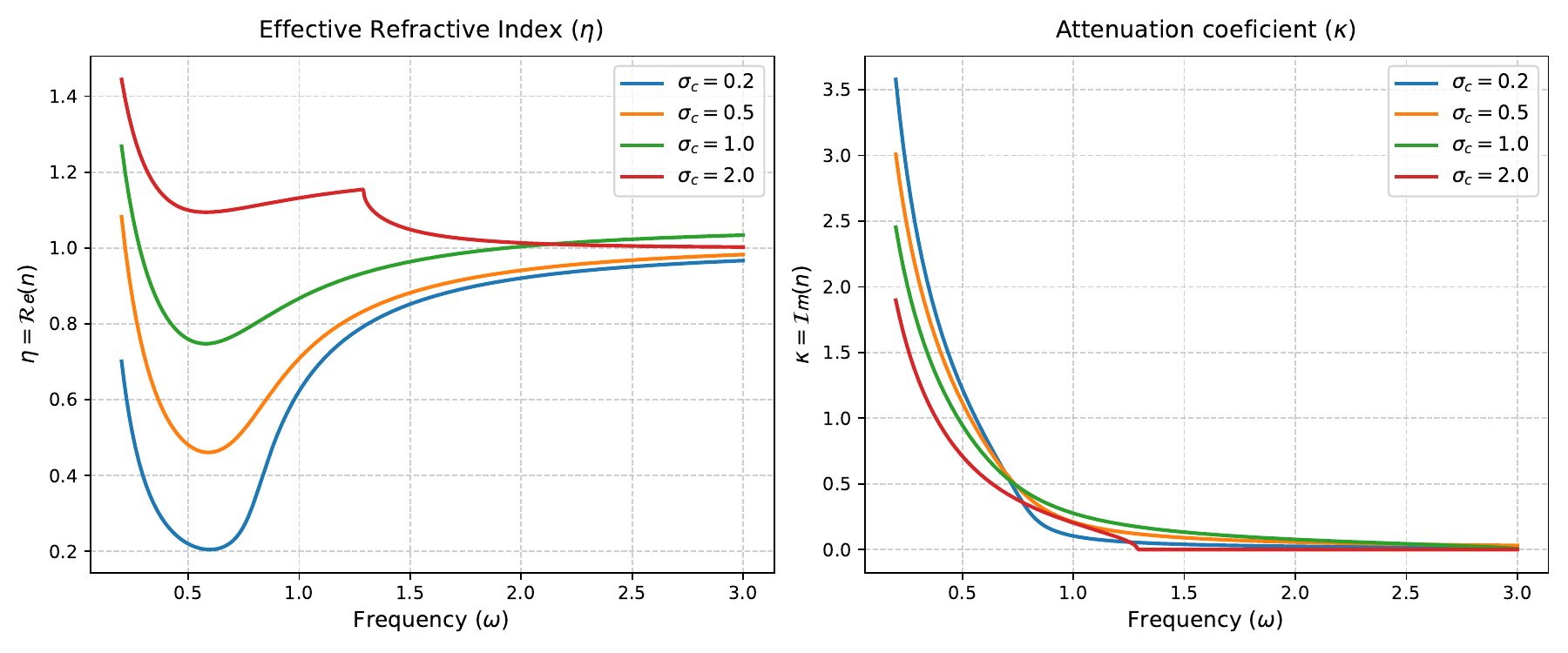}
    \caption{Left panel: The real part $(\eta)$ of the refractive index as a function of the $\omega$-frequency, for the physical root of the system ($m=0$), for the values of electric conductivity $\sigma_c=0.2$, $\sigma_{c}=0.5$, $\sigma_{c}=1.0$ and $\sigma_{c}=2.0$. Right panel: The imaginary part $(\kappa)$ for the refractive index as a function of the $\omega$-frequency, for the solution of $m=0$. We use the same values of $\sigma_{c}$ from the left panel.}
    \label{fig:dispersao_exata}
\end{figure*} 
\begin{widetext}
\begin{eqnarray}\label{eq:n_complexo}
n_m(\omega) = \Bigg\{ 1 - \Bigg[ \, \xi^m \Bigg( -\frac{i}{2} \left[ \frac{2\theta^2}{\sigma_c\omega^2} + \frac{4-\sigma_c^2}{6\sigma_c} + \frac{(4-\sigma_c^2)^3}{108\sigma_c^3} \right] 
 \nonumber \\
+\sqrt{ -\frac{1}{4}\left[ \frac{2\theta^2}{\sigma_c\omega^2} + \frac{4-\sigma_c^2}{6\sigma_c} + \frac{(4-\sigma_c^2)^3}{108\sigma_c^3} \right]^2 + \left[ \frac{1}{3} + \frac{(4-\sigma_c^2)^2}{36\sigma_c^2} \right]^3 } \;\Bigg)^{\frac{1}{3}} 
\nonumber \\
+ \xi^{3-m} \Bigg( -\frac{i}{2} \left[ \frac{2\theta^2}{\sigma_c\omega^2} + \frac{4-\sigma_c^2}{6\sigma_c} + \frac{(4-\sigma_c^2)^3}{108\sigma_c^3} \right] 
\nonumber \\
-\sqrt{ -\frac{1}{4}\left[ \frac{2\theta^2}{\sigma_c\omega^2} + \frac{4-\sigma_c^2}{6\sigma_c} + \frac{(4-\sigma_c^2)^3}{108\sigma_c^3} \right]^2 + \left[ \frac{1}{3} + \frac{(4-\sigma_c^2)^2}{36\sigma_c^2} \right]^3 } \;\Bigg)^{\frac{1}{3}} 
+ i\left(\frac{4 - \sigma_c^2}{6\sigma_c}\right) \, \Bigg]^2 \, \Bigg\}^{\frac{1}{2}} \; .
\end{eqnarray}
\end{widetext}
%
%

This result shows that the non-local dispersion problem, intrinsic to the non-local operator of the model, is strictly mapable to algebraically obtaining the roots of a third-degree polynomial. This formulation eliminates the need for perturbative expansions in the long-wavelength limit. The refractive index $n_{m} \, (m=0,1,2)$, obtained as an intrinsically complex quantity, already exactly encapsulates all the physical information of the system, simultaneously determining the propagation, and the Ohmic attenuation characteristics of the electromagnetic wave, bypassing the need for an explicit algebraic separation.


The consistence of this algebraic formulation can be attested by analyzing the asymptotic limits, which fully recover the phenomenology of both the purely resistivity, and the purely topological regimes. 
%
%
%
%
We plot numerically the real and imaginary parts of (\ref{eq:n_complexo}) as functions of the wave frequency for some values of electric conductivity in the figures 2 and 3, respectively.    
%

%
The analysis of the imaginary spectrum provides the wave attenuation ($\kappa$) is severe in the high-frequency regime, 
in agreement with the topological mass \textit{gap}. The real spectrum increases in the dissipative range as the conducting lattice substantially grows imposing a drastic retardation on the phase velocity of the radiation. In particular, in the strong-conductivity regime ({\it e.g.}, $\sigma_c = 2.0$), a derivative singularity emerges in the real dispersion, highlighting a phase transition between the underdamped and overdamped regimes of the system.
\subsection{The high-conductivity limit and shielding in graphene}
Although Eq. (\ref{eq:n_complexo}) includes the complete phenomenology of non-local damping, it is of technological interest to investigate the asymptotic limit for strong conductors. For highly conducting two-dimensional materials, such as, the heavily doped graphene, the physical regime is $\sigma_c \gg 1$.  Under this approximation, the analysis of the root associated with the hybridized TM mode in Eq. (\ref{zm}) reveals that the dynamics becomes governed exclusively by the balance between transverse polarization and dissipation. In this limit, the topological term becomes analytically subdominant, and the root of interest collapses to the asymptotic form, {\it i. e.}, $z_{TM} \approx - \, i \, \sigma_c/2$.

By reverting the parametrization through the relation $n = \sqrt{1 - z^2}$, the real and imaginary components of the plasmon's refractive index strictly decouple : $\eta(\omega) \approx \frac{\sigma_c}{2}$, and $\kappa(\omega) \approx 0$.

This limit has a profound characteristic of the system: at extreme conductivities, the electric field tangent to the planar sheet is perfectly shielded ($E_{\parallel} \rightarrow 0$). Consequently, there is no transfer of mechanical work to the crystal lattice, which extinguishes the Joule heating ($\kappa \approx 0$). Simultaneously, the wave is subjected to an extreme confinement, resulting in an effective refractive index $(\eta)$ that runs linearly with the sheet's conductivity. Thus, the wave propagates as a purely subluminal and lossless mode, with a phase velocity of $v_p \approx 2c/\sigma_c$. 
The figure 3 corroborates this analytical limit, showing the scaling of the effective refractive index, and the suppression of the dissipative spectrum at high conductivities.
\section{Conclusions}
In this work, we have investigated the wave propagation, and the hybridization of surface plasmon-polaritons described by the pseudo-electrodynamics added to a non-local Chern-Simons topological term. We derived the exact wave equations for the observable fields and the gauge-potential, demonstrating that the topological parameter $\theta$ introduces an energy \textit{gap} and it breaks the strict transversality of the waves by generating a substantial longitudinal electric component.
The fundamental importance of these results lies in the revelation that the intrinsic topology of planar space-time acts as a mechanism for a dynamical mass generation and shielding against low frequencies, without the need to couple additional scalar fields. Furthermore, we have showed the analytic solution for the refractive index for a planar material governed by the Ohm law in which the topological mass drives a hybridization of the TE- and TM-modes. This phenomenon emulates an intrinsic Hall Effect that forms a chiral and elliptically polarized surface plasmon, topologically protected against the backscattering.
From an experimental perspective, the observation of this topologically mediated TE-TM hybridization demands specific material conditions to prevent the non-local effects from being suppressed. The surface conductivity cannot operate in the strict limit of a perfect conductor ($\sigma_c \gg 1$), where the electric field is trivially shielded. Instead, it requires moderately doped, high-mobility samples—such as graphene encapsulated in hexagonal boron nitride (hBN)—to minimize the dissipative scattering rate ($\gamma$) so that the topological mass scale dominates ($\theta > \gamma$). Probing such devices within the Terahertz frequency window would allow the dynamic conductivity to sustain the plasmonic mode without obliterating the non-local topological signature.
In summary, the formulations established the robustness of the PED-MCS as an effective gauge theory. These results strengthen the mathematical foundation of non-local electrodynamics in $1+2$ dimensions and provide a theoretical framework for future investigations in condensed matter physics. The perspective 
opened by this work are directed towards the study of macroscopic interactions in topological insulators, the phenomenology of novel planar semiconductor, and superconductor devices focused on advanced \textit{hardware} applications, where the topological protection of plasmonic modes will play a central role.

\end{document}